\documentstyle[11pt]{article}
\title{\begin{flushright}
{\normalsize NUC-MINN-94/5-T\\
August 1994 \\}
\end{flushright}
\bf Nucleation of Quark--Gluon Plasma from Hadronic Matter}
\author{{\bf Joseph I. Kapusta} and {\bf Axel P. Vischer}\\
  {\small\it School of Physics and Astronomy,
   University of Minnesota, Minneapolis, MN 55455}
\and
{\bf Raju Venugopalan}$^*$\\
   {\small\it Theoretical Physics Institute,
   University of Minnesota, Minneapolis, MN 55455}}

\date{}

\begin{document}

\maketitle

\begin{center}
Abstract
\end{center}

\noindent
The energy densities achieved during central collisions of large nuclei
at Brookhaven's AGS may be high enough to allow the formation of
quark--gluon plasma.  Calculations based on relativistic nucleation
theory suggest that rare events, perhaps one in every 10$^2$ or
10$^3$, undergo the phase transition.  Experimental ramifications
may include an enhancement in the ratio of pions to baryons, a
reduction in the ratio of deuterons to protons, and
a larger source size as seen by hadron interferometry.

\vspace*{1.0in}

\vspace*{1.0 in}
\noindent
$^*${\small Address after Sept. 1, 1994:
{\it Institute for Nuclear Theory, University of Washington, Seattle,
WA 98195.}}

\vfill \eject

\section{Introduction}

Experiments at Brookhaven's AGS with beams of oxygen, sulphur and gold
at laboratory energies of 10 to 15 GeV per nucleon have indicated a
nearly complete stopping of the nuclei during central collisions
\cite{qm}.  This massive pile-up of nuclear matter is also
seen in numerical simulations which approximate the nuclear collisions
as a sequence of elementary hadron-hadron collisions, such as RQMD \cite{rqmd}
and ARC \cite{arc}.  Energy densities of up to 2 GeV/fm$^3$ may be
realized in the laboratory. One may legitimately ask the question:
Is quark--gluon plasma produced during these collisions?  Despite the
fact that most experimental data so far are consistent with the
hadron-based cascade simulations we suggest
that the answer may be yes, at least in rare events.

The basic picture we have in mind is as follows.  During the initial stage
of the collision the nuclei stop each other and get heated due to
elementary nucleon-nucleon collisions and the associated production
of mesons.  Occasionally the local energy density may reach a very
high value due to fluctuations.  In this small region of space the
matter is more readily described as a plasma droplet of quarks and gluons
rather than as a gas of hadrons.  If the average energy density in
the space surrounding this plasma droplet is above a certain critical
value then the plasma droplet will grow, converting more hadrons to
quark--gluon plasma.  Since there are no containment walls the matter,
whatever phase it is in, will eventually expand and cool.  In the
end all quarks and gluons must be rehadronized and will be detected
as such.

Quantitative questions now arise.  Is the energy density achieved at
the AGS high enough?  How big must a plasma droplet be to grow?
What is the time scale for producing such a critical size droplet?
How much of the total volume is converted to plasma, and how long
does it last?  Of course, in order to ask these questions we must
assume the existence of a deconfinement/chiral symmetry restoring
phase transition, or at least a rapid crossover.

The picture at the AGS is different than that expected at Brookhaven's
RHIC which is now under construction.  At RHIC, where the energy
is to be 100 GeV per nucleon per beam, the nuclei are expected to be
transparent to each other.  Hard collisions between and among the
quarks and gluons in the nuclear structure functions will produce
a hot, nearly baryon-free, plasma in the central rapidity region,
the so-called inside-outside cascade \cite{BJ,Klaus}.  The receding
nuclei will be compressed and heated \cite{Larry}.  As the matter
expands and cools it will undergo a hadronization phase transition
as bubbles of hadronic matter are nucleated in the pre-existing
quark--gluon plasma \cite{Csernai}.  The picture is the same for
lead on lead collisions at
CERN's anticipated LHC.  The situation at CERN's existing SPS
is not clear.  At its lower energies it may be like the AGS, and
at its maximum energies of 100 to 200 GeV per nucleon in the
laboratory frame it may be more similar to RHIC.

The approach we follow is analogous to that taken under the
assumption that the transition begins in the quark--gluon plasma phase,
as appropriate for RHIC \cite{Csernai}.  First, we describe a very simple
model parametrizing the time evolution of the hadronic matter
in a central collision assuming complete stopping.  Second, we
convert the baryon and energy densities into temperatures and
chemical potentials via the use of a hadronic equation of state.
We also need an equation of state describing the quark--gluon
plasma phase.  Third, we determine the rate of nucleation of plasma
droplets in superheated hadronic matter and their subsequent growth
velocities.  Then we put it all together and solve the resulting
equations numerically.  Those interested only in the results
may turn directly to section 5.

We discuss and propose some experimental signatures in the conclusion.

\section{Dynamics of Nuclear Collisions}

The dynamics of a central nucleus-nucleus collision at the AGS is
extremely complicated.  We shall be satisfied with a simple model for
an exploratory excursion into the problem of nucleation
of plasma.  To first approximation this model is consistent with
the ARC cascade simulations and with direct experimental measurements.

Imagine the colliding nuclei as two Lorentz contracted disks in the
center of momentum frame.  At time $t = 0$ they touch.  They
interpenetrate between $0 \leq t \leq t_0$ where $t_0 = R/\gamma$,
$R$ is the nuclear radius, and $\gamma$ is the Lorentz factor in
the center of momentum frame.  At the end of this time the nuclei
are completely stopped.  The volume of overlap as a function of time is
\begin{eqnarray}
V(t) \, = \, \frac{V_0 \, t}{t_0} \;\;\;\;\;\;\;\;\; 0 \leq t \leq t_0 \, ,
\label{volume}
\end{eqnarray}
where $V_0 = 4\pi R^3/3$.  The matter within this overlap volume
is assumed to be thermalized with constant baryon density
$2 \gamma n_0$ and energy density $2 \gamma^2 m_N n_0$, where
$n_0$ is normal nuclear matter density and $m_N$ is the nucleon
mass.

After the time $t_0$ the hot fireball expands radially.  At late
times we would expect its radius to grow linearly with time.
Therefore we parametrize the volume as $V(t) = A (t + a)^3$.
The constants $A$ and $a$ are determined by matching the volume
and its first derivative at $t_0$.  This gives for the expansion
volume
\begin{eqnarray}
V(t) \, = \, V_0 \left( \frac{t + 2t_0}{3t_0} \right)^3
\;\;\;\;\;\;\;\;\; t_0 \leq t \, .
\label{also-v}
\end{eqnarray}
Eventually the particles will begin free-streaming, but we shall not
be interested in what happens at such low densities.

The time dependence of the baryon density is
\begin{eqnarray}
{n_B(t) \, = \, \left\{ \begin{array}{ll}
2 \gamma n_0 & t \leq t_0 \\
2 \gamma n_0 \left[3t_0/(t + 2t_0) \right]^3 & t_0 \leq t \, .
\end{array} \right. }
\label{n-density}
\end{eqnarray}
We assume an entropy-conserving hydrodynamic expansion.  Hence the
entropy density is
\begin{eqnarray}
{s(t) \, = \, \left\{ \begin{array}{ll}
s(t_0) & t \leq t_0 \\
s(t_0) \left[3t_0/(t + 2t_0) \right]^3 & t_0 \leq t \, .
\end{array} \right. }
\label{s-density}
\end{eqnarray}
In other words, the entropy per baryon is constant during the expansion.
The initial entropy density $s(t_0)$ must be determined from the
initial baryon and energy densities via an equation of state.

To get some typical numbers consider gold on gold collisions
at a beam energy of 11.6 GeV per nucleon.  Then $R = 7$ fm and
$\gamma = 2.7$ resulting in a characteristic time of 2.6 fm/c,
an initial baryon density of 0.78 fm$^{-3}$ and an initial energy
density of 1.95 GeV/fm$^3$.  These numbers are very similar to those
obtained from ARC with the caveat that the matter is not completely
thermalized in the cascade simulation.

\section{Equation of State for Baryon Rich Matter}

In this section we discuss the equation of state for the quark--gluon
plasma and the hadron gas. Despite much progress in lattice QCD studies
there is much uncertainity in the equation of state results when
dynamical quarks are included.  In addition, current lattice results
provide little insight into the equation of state for the large baryon
chemical potentials relevant to this work.  We shall therefore
(i) assume that the hadron to quark--gluon phase transition is
first order, (ii) use simple models to describe the equation of state
in each of the two phases, and (iii) perform a Maxwell construction
to join the two phases along their common boundary.

For simplicity we will assume that the quark--gluon plasma consists
of a free gas of quarks and gluons with a bag
constant to represent confinement.  For a plasma with up, down and
strange quarks we choose the independent variables to be the temperature
$T$, the chemical potential for up and down quarks $\mu_u = \mu_d
= \mu_q$ under the assumption of charge symmetric matter \cite{note1},
and the chemical potential for strange quarks $\mu_s$. Since the
strong interactions conserve strangeness, and there is insufficient
time for the weak interactions to be operative, the plasma has no
net strangeness; this requirement implies that $\mu_s=0$.  We collect
below the expressions for the pressure, the baryon density, the entropy
density and the energy density in the quark--gluon phase:
\begin{eqnarray}
P_{\rm qg}&=& {{32 + 42 + 21 f_1(m_s/T)}\over 180}\pi^2 T^4 + \mu_q^2 T^2 +
{1\over {2\pi^2}}\mu_q^4 - B \, ,\nonumber \\
n_B^{\rm qg}&=& {2\over 3} \mu_q \left( T^2 + {1\over \pi^2}\mu_q^2
\right) \, , \nonumber \\
s_{\rm qg}&=&{{32 + 42 + 21 f_2(m_s/T)}\over 45}\pi^2 T^3 +
2T\mu_q^2 \, ,\nonumber \\
\varepsilon_{\rm qg}&=&-P_{\rm qg}+Ts_{\rm qg}+3\mu_q n_B^{\rm qg} \, .
\end{eqnarray}
The baryon chemical potential is $\mu_B=3\mu_q$, the electric charge
chemical potential is $\mu_Q=0$, and the strangeness chemical
potential is $\mu_S=-\mu_q$.  The bag constant $B$ is chosen to
be (220 MeV)$^4$.  The up and down quark masses are set to zero.
The strange quark mass, $m_s$, is somewhere in the range of 150 to
280 MeV.  The functions $f_1(m_s/T)$ and $f_2(m_s/T)$ involve
a momentum integration over the strange quark Fermi distribution function.
Their limits are $f_1(0) = f_2(0) = 1$ and $f_1(\infty) = f_2(\infty)
=0$.  Rather than doing the integrals
numerically, we will ignore the strange quark in the plasma phase
altogether.  The matter at the AGS is quite baryon rich, and the
chemical potential $\mu_q$ is typically of order 500 MeV.  Since the
temperature is typically of order 200 MeV, the strange quark contributes
very little.  As an example, if the strange quark is included with
zero mass then the critical temperature at zero baryon density
is about 150 MeV.  If the strange quark is not included, but nothing
else is changed, then the critical temperature is 161 MeV.  A realistic
quark mass would give something in between.  For increasing baryon
density the difference gets even smaller.

For the hadronic equation of state we consider a gas of mesons
($\pi$, $K$, $K^{*}$, $\eta$, $\eta^\prime$, $\rho$, $\omega$,
$\phi$ and $a_1$) and baryons (nucleons, $\Delta$, $\Lambda$ and
$\Sigma$) and the corresponding anti-baryons.  The only interaction
we directly account for is the repulsive mean field in the baryon
sector.  This is done in the usual way by adding a term proportional
to the baryon density to the baryon chemical potential \cite{K}.
\begin{equation}
\mu_B^* \, = \, \mu_B - Kn_B \, .
\end{equation}
Here $K$ is the strength of the repulsive mean field.  The chemical
potential for a hadron of type $i$ is expressed in terms of the
baryon, electric charge, and strangeness chemical potentials as
\begin{equation}
\mu_i \, = \, B_i \, \mu_B^* + Q_i\,\mu_Q + S_i\,\mu_S \, ,
\end{equation}
where $B_i$, $Q_i$ and $S_i$ are the corresponding quantum numbers
of the hadron.

The pressure in the hadronic phase is \cite{K}
\begin{eqnarray}
\lefteqn{P_{\rm had} \, = \, \frac{1}{2}Kn_B^2 \, + } \nonumber \\
& & T\sum_i g_i \int {d^3q \over {(2\pi)}^3} \bigg[
\ln\left(1 \pm e^{-\beta(\epsilon_i-\mu_i)}\right)^{\pm 1}+
\ln\left(1 \pm e^{-\beta(\epsilon_i+\mu_i)}\right)^{\pm 1} \bigg] \, ,
\end{eqnarray}
where $\epsilon_i = \sqrt{q^2+m_i^2}$ and the $\pm$ refer to fermions
or bosons.  The baryon density, electric
charge density, and strangeness density are all determined in the
usual way by differentiating the pressure with respect to $\mu_B$,
$\mu_Q$ and $\mu_S$, respectively.

The baryon density must be determined self-consistently by solving the
nonlinear equation
\begin{equation}
n_B \,=\, \sum_{i=N,\Delta,\Lambda,\Sigma} g_i
\int {d^3q \over (2\pi)^3}
\left\{ {1\over {\exp\left[\beta(\epsilon_i-\mu_i) \right]+1}}-
{1\over {\exp\left[ \beta(\epsilon_i+\mu_i)\right]+1}}
\right\} \, .
\end{equation}
Since we require the net strangeness of the hadrons to be zero
and the charge to baryon ratio to be 1/2 these additional constraints
must be implemented simultaneously with the one for the baryon density.

The final step is to perform the Maxwell construction joining the two
phases.  This is done by equating the temperatures, baryon chemical
potentials and pressures in the two phases. In the top panel of
figure 1 we plot the coexistence curve in the $T$--$\mu_B$ plane
for a bag constant $B$ = (220 MeV)$^4$ and a mean field parameter
$K=1500$ MeV$\cdot$fm$^3$.  For each temperature/chemical potential
point on the coexistence curve there is a particular value of the
pressure.  Since the energy density, number density and entropy density
are all discontinuous across the coexistence curve, there is a
mixed phase consisting of different regions of space which are
in either the quark--gluon or the hadron phase.  This is illustrated
in the bottom panel of figure 1 for the $T-n_B$ plane.  To interpolate
across the boundary one introduces the quark--gluon fraction $q$
which ranges from 0 to 1.  The hadron fraction is then $1-q$.
\begin{eqnarray}
\varepsilon_{\rm mix} &=& (1-q) \,\varepsilon_{\rm had} +
q \, \varepsilon_{\rm qg} \, ,\nonumber \\
n_B^{\rm mix} &=& (1-q) \, n_B^{\rm had} +
q \, n_B^{\rm qg} \, ,\nonumber \\
s_{\rm mix} &=& (1-q) \, s_{\rm had} + q \, s_{\rm qg} \, .
\end{eqnarray}

It is worth remarking that without the incorporation of a bag constant
$B$ or a repulsive baryon mean field $K$ one generally does not
get a sensible transition from hadrons to quarks and gluons in the
whole $T-\mu_B$ plane.

\section{Nucleation Rate for Baryon Rich Matter}

The rate $I$ to nucleate droplets of quark--gluon plasma in a hadronic
gas per unit time per unit volume is given by \cite{Lang1,GunMiSa,JL1}
\begin{eqnarray}
I=\frac{\kappa}{2\pi} \Omega_0 \exp(-\Delta F_*/T) \, .
\label{eqt1}
\end{eqnarray}
Here $\kappa$ is the dynamical prefactor, $\Omega_0$ is
the statistical prefactor, and $\Delta F_*$ is the change in free energy
of the system due to the formation of a single critical size droplet of
plasma.  Each of the three factors will be discussed in turn.

The nucleation process is driven by statistical fluctuations which
produce droplets of quark--gluon plasma in the hadronic phase.
The size of these fluctuations is determined by the free energy difference of
the hadronic phase with and without the plasma droplet. This energy
difference can be approximated by a liquid-drop expansion~\cite{expansion}
\begin{eqnarray}
\Delta F = \frac{4 \pi}{3}R^3 \left[ P_{\rm had}(T,\mu_B)
-P_{\rm qg}(T,\mu_B) \right] +4 \pi R^2 \sigma
+\tau_{\rm crit} T \ln \left[1+\frac{4 \pi}{3} R^3 s_{\rm qg} \right] \, .
\label{delf}
\end{eqnarray}
The first term represents the usual volume or pressure contribution,
the second term is the surface contribution which is proportional to
the surface tension $\sigma$,
and the third term is the so-called shape contribution. Close to the phase
transition the volume contribution approaches zero. The shape
contribution is an entropy term which takes into account small
fluctuations in the shape of the droplet which conserve both the
volume and the surface area (Fisher's magic carpet effect).  It is
proportional to the logarithm of the entropy of the quark--gluon droplet
$\frac{4 \pi}{3} R^3 s_{\rm qg}$. The 1 under the logarithm is added to
ensure regular behavior at $R \rightarrow 0$. The critical exponent
$\tau_{\rm crit}$ would determine the behaviour of the distribution
close to a critical point where the surface tension vanishes and
the phase transition is second order.  Such a critical point exists
for liquid-gas type of phase transitions, but does not exist in the
scenario of the hadron to quark-gluon phase transition assumed here.
Generally $\tau_{\rm crit}$ is slightly larger than 2.  Figure 2 shows
a sketch of $\Delta F$ as a function of $R$.

The system under discussion is in a superheated state so that the pressure
difference in equation (\ref{delf}) is negative. Minimizing $\Delta F$ with
respect to the droplet radius $R$ yields the critical radius $R_*(T,\mu_B)$.
Droplets with a radius larger than $R_*$ will expand into the hadronic phase,
while droplets with a radius smaller than $R_*$ will collapse.  $\Delta F_*$
is the activation energy needed to create a droplet of critical size $R_*$.

The dynamical prefactor $\kappa$ determines the exponential growth rate of
critical-size droplets.  For the droplets to grow beyond the critical
radius, latent heat must be carried to the surface of the droplet
from the surrounding hadronic matter. This
is achieved through thermal dissipation and/or viscous damping.
The general result for the dynamical prefactor is \cite{RAJ}
\begin{eqnarray}
\kappa = \frac{2\sigma}{(\Delta w)^2 R_*^3} \left[\lambda
T+2\left(\frac{4}{3}\eta+ \zeta\right)\right] \, .
\end{eqnarray}
Here $\Delta w$ is the difference in enthalpy densities of the two phases.
$\lambda$ is the thermal conductivity and $\eta$ and $\zeta$ are the
viscosities of the hadronic phase.  Notice that $\kappa$ is linearly
proportional to the dissipative coefficients, as expected for
linear viscous fluid dynamics.

For the dissipative coefficients we use the parametrization of Danielewicz
\cite{Pavel}, extrapolated to the region of temperatures and baryon
densities we are interested in.
\begin{eqnarray}
\eta &=&\left(\frac{1700}{T}\right)^2 \left(\frac{n}{n_0}\right)^2
 + \frac{22}{1+T^2/1000} \left(\frac{n}{n_0}\right)^{0.7}
+ \frac{5.8 \, T^{1/2}}{1 + 160/T^2} \, ,\\
\lambda &=& \frac{0.15}{T} \left(\frac{n}{n_0}\right)^{1.4}
+ \frac{0.02}{1 + T^4 / 7 \times 10^6} \left(\frac{n}{n_0}\right)^{0.4}
+ \frac{0.0225 \, T^{1/2}}{1 + 160/T^2} \, .
\end{eqnarray}
Here $T$ is given in MeV, $\eta$ in Mev/fm$^2$c and $\lambda$ in c/fm$^2$.
The bulk viscosity $\zeta$ is neglected since it is a lot smaller than
the shear viscosity $\eta$.

The free energy $\Delta F$ given in (\ref{delf}) is a functional of a set of
collective variables, chosen here to be the local temperature $T$ and
the chemical potential $\mu_B$.  Using the equations of state we could have
chosen instead the local energy density $\varepsilon$ and baryon density
$n_B$.  Figure 2 is a one-dimensional
projection of this space of collective variables. The statistical prefactor
$\Omega_0$ is a measure of the phase space volume of the saddle point region of
the free energy functional and $\Delta F_*$ is the change in the free energy
required to cross the saddle.
To first approximation the statistical prefactor is
\cite{Lang1,Lang3,JL1,RAJ}
\begin{eqnarray}
\Omega_0 = \frac{2}{3 \sqrt{3}} \left( \frac{\sigma}{T} \right)^{3/2}
\left( \frac{R_*}{\xi_{\rm had}} \right)^4 .
\label{omega}
\end{eqnarray}
The correlation length in the hadronic phase, $\xi_{\rm had}$,
is expected to be on the order of 0.5 to 1.0 fm at the relevant energy
densities.  Higher order corrections to $\Omega_0$, arising from
fluctuations, are already included phenomenologically in $\Delta F$
when we evaluate it with the {\it measured} values of the surface
tension, equation of state, and shape contribution.  See Langer
and Turski \cite{LT73}.

There are several crucial assumptions inherent in this expression for
the nucleation rate. First, it is assumed that the phase transition
is of first order.  Second, it is assumed that the temperature and
chemical potentials are well defined, and vary more or less smoothly
and slowly throughout the system.  Third, it is assumed
that when nucleation takes place the critical-size droplet has a
radius which is no smaller than the correlation length, otherwise
the validity of statistical averaging becomes dubious.  We do
not necessarily believe that these assumptions are correct in detail;
rather, we use them as a basis to present interesting possibilities.

To be concrete in what follows, we take the surface tension to
be $\sigma$ = 50 MeV/fm$^2$, the correlation length in the hadronic
phase to be $\xi_{\rm had}$ = 0.7 fm, and the critical coefficient
to be $\tau_{\rm crit}$ = 2.2.  In principle $\sigma$ cannot be
varied completely independently of the equation of state.  If the
latent heat goes to zero so that the first order phase transition
goes over into a second order one, $\sigma$ must go to zero also.
Conversely, as the latent heat increases, one might expect $\sigma$
to become larger.  Lattice gauge theory calculations so far give us no
information on its magnitude at large baryon densities.

The expressions given here for the various components of the nucleation
prefactor are relevant for one dense phase of matter (quark-gluon plasma)
immersed in another (hadronic matter).  The basic physics is that initial
droplet (bubble) growth is limited by the ability of viscosity and heat
conduction to carry latent heat to (away) from the surface.  One may consider
a different scenario where a dense droplet of one phase (quark-gluon plasma)
is surrounded by a dilute gas of the other phase (hadronic matter).
The plasma droplet then would grow by accretion of individual hadrons.
Although we don't think that this is the relevant situation, since
in the superheated hadronic phase the mean free path and correlation
length are of order 1 fm or less, we include an appendix outlining
the theoretical expression for the prefactor if it were.

Nucleation begins when the two nuclei first collide with each other
and a superheated overlap region is created. It ends when the
expanding system reaches the phase coexistence curve. The fraction of
space which is converted into quark--gluon matter $q$ is computed
from the expression \cite{Csernai}
\begin{equation}
q(t) =  \frac{1}{V} \int^{t}_{0} dt' I(t') V(t')
[1-q(t')] V_{\rm drop}(t,t') \, .
\label{qu}
\end{equation}
The total volume of the system at time $t$ is $V(t)$.
The volume already occupied by quark--gluon plasma is not available
for nucleation.  The volume which is available for nucleation is
$V(t)[1-q(t)]$.  Once a drop has been nucleated, with
radius $R_* (T,\mu_B)$, it will grow radially with speed $v(T,\mu_B)$.
The volume of the droplet is therefore a nonlocal function of time.
It can be written as
\begin{eqnarray}
V_{\rm drop}(t,t') = \frac{4 \pi}{3} \left[ R_*(t') + \int^{t}_{t'} dt''
\, v(t'') \right]^3 \, .
\label{vdrop}
\end{eqnarray}
$I$, $R_*$ and $v$ all depend on time because they depend on the
(time dependent) temperature and chemical potential.

The speed $v(t)$ with which the droplet expands into the hadronic
matter is relatively unknown. To determine this speed would require
a detailed microscopic study of the system. Instead we make the
plausible assumption that the expansion into the new phase is driven
by the pressure difference $\Delta P(t) = P_{\rm qg}(t)-P_{\rm had}(t)$
between them \cite{wall-speed}.  The greater the pressure difference
the faster the plasma droplet expands.  As the critical curve is
approached the pressure difference goes to zero, and so should the
droplet expansion velocity, since on the critical curve neither of
the phases is thermodynamically preferred over the other.
Thus we write
\begin{equation}
\gamma v = v_0 \frac{\Delta P}{P_{\rm qg}} \, ,
\label{speed}
\end{equation}
where $\gamma = 1/\sqrt{1-v^2}$; $v_0$ is a free phenomenological
parameter which we expect to be on the order of 1.

\section{Numerical Results}

We consider two sets of initial conditions.  Set 1 corresponds to a
central gold - gold collision with a lab kinetic energy
of $E_{\rm kin}$ = 11.6 GeV/A and $\gamma$ = 2.68, as achieved at
Brookhaven's AGS.  Set 2 corresponds to a central
lead - lead collision with a lab kinetic energy of $E_{\rm kin}$
= 25 GeV/A and $\gamma$ = 3.78, as could be achieved at CERN's SPS.
The average baryon density in a nucleus is taken as $n_0$ = 0.145 fm$^{-3}$
so that the radii of the nuclei are given by $R = r_0 A^{1/3}$ with
$r_0$ = 1.18 fm.  The interpenetration time $t_0 =
R/ \gamma$, defined in (\ref{volume}), is  2.56 fm (1.85 fm), and the initial
baryon density in the overlap region, defined in (\ref{n-density}), is
0.78 fm$^{-3}$ (1.10 fm$^{-3}$) for set 1 (2), respectively. The energy density
$2 \gamma^2 m_N n_0$ reached in this first stage of the collision is thus
1.95 GeV/fm$^3$ (3.90 GeV/fm$^3$) for set 1 (2).  Table 1 gives a
summary of the initial conditions for the two parameter sets.  Some
of the quantites in this table, like the temperature and baryon chemical
potential, are dependent on the equation of state, while others are not.

The assumed volume of kinetically equilibrated matter $V(t)$ is
determined by equations (\ref{volume}) and (\ref{also-v}).
Figure 3 displays the volume as a function of
time for the two parameter sets.  We see the linear increase of the overlap
volume up to the interpenetration time $t_0$ after which the nuclei are
completely stopped and start to expand spherically, leading to a cubic
increase with time.

Knowledge of the time evolution of the volume allows us to evaluate the
time dependence of the baryon and entropy densites, as given in equations
(\ref{n-density}) and (\ref{s-density}). With the help of the equations
of state discussed in section 3 we can then evaluate the chemical potential
$\mu_B(t)$ and the temperature $T(t)$ of the hadronic phase as functions of
time.  Figure 4 shows them as well as the baryon density and the
energy density for both parameter sets. The ordinates are normalized
to their initial values as displayed in Table 1.

Next we plot the path of the collision in the $\mu_B$--$T$ plane
of figure 1. The system starts out as an extremely superheated
hadron gas deep within what ought to be the quark--gluon phase at a
temperature of 173 MeV (214 MeV) and a chemical potential of
1724 MeV (2064 MeV). It stays at this point
up to the interpenetration time $t_0$, then expands and cools,
reaching the phase coexistence curve at a time $t_f$ = 7 fm (6.95 fm)
later.

It is important to recognize that we are neglecting the feedback
of quark-gluon plasma nucleation on the temporal evolution of
the temperature and chemical potential.  We shall return to this
point later.

In figure 5 we plot the nucleation rate $I(t)$ along the path
in the $T-\mu_B$ plane for the two parameter sets.
During interpenetration both temperature and chemical potential are
constant; the nucleation rate is therefore also constant.
After $t_0$ the rate first increases, reaches a maximum, then
decreases to zero as the coexistence curve is approached.
We would expect the rate of nucleation of plasma to {\it increase}
as the initial state of the system gets further from the
phase coexistence curve, and therefore the rate should be a monotonically
{\it decreasing} function of time after $t_0$.  Why isn't it?
The reason is rather fundamental.  The usual analytic expression for
the nucleation rate, equation (11), is derived under the assumption
that the system has been either superheated or supercooled just a
small amount from the phase coexistence curve.  This means that
the trajectory in phase space which goes from a metastable point
to a stable point is dominated by a saddle, and the saddle
configuration is a spherical droplet or bubble.  All other
configurations have a $\Delta F/T$ which is significantly larger,
and therefore exponentially suppressed in comparison.  When
the system is superheated as dramatically as it is in our examples,
$\Delta F_*/T \approx 2$.  Then the dominant contribution to $\Delta F_*$
is from the shape term.  More extreme configurations with shapes
like lasagna and spaghetti ought to be contributing too.  However, these
are difficult to take into account in any simple manner, especially
concerning the preexponential factors.

Mathematically, the reason
the rate turns over in this figure can be explained this way:
Going away from the phase coexistence curve, the Boltzmann exponential
increases.  The preexponential factor is proportional to
\begin{displaymath}
\frac{R_*}{T^{3/2}(\Delta w)^2} \left[\lambda
T+\frac{8}{3}\eta\right] \, .
\end{displaymath}
The dissipation coefficients are relatively slowly varying functions,
and increase by only 20 or 30\%.  But $R_*$ gets smaller, $T$ gets
bigger, and the enthalpy difference squared increases by about an
order of magnitude.  Overall, the preexponential factor decreases.
Multiplication of a decreasing function by an increasing function results,
in this case, in a product with a maximum, as seen in figure 5.
Actually, the dominant effect comes from $\Delta w$, and this arises
from the extreme superheating of hadronic matter in the collisions.
It is quite likely that we are {\it underestimating} the initial
nucleation rate by an order of magnitude.  In reality, for small
droplets $\sigma$ effectively depends on $R$ \cite{Cahn,Scriven}.
With increasing superheating one eventually reaches a point where
spinodal decomposition sets in and the phase transition will be
extremely rapid \cite{Cahn}.
This situation is not very well understood.  Still, an uncertainty
even this large is acceptable in a first study of this nature.

In figure 6 we plot the average droplet density, defined by
\begin{equation}
n_{\rm drop} (t) \,=\, \int^{t}_{0} dt' \, I(t') \, ,
\end{equation}
as a function of time.  The average droplet density is independent
of the droplet growth speed as long as $q(t)$ remains small, which
it does in these examples.  The maximum possible value reached by the
droplet density is $n_{\rm drop} (t_f)$ = 2 (1) $\times 10^{-5}$
fm$^{-3}$ for parameter set 1 (2). The volume of
the expanding system, on the other hand, is on the order of
2 $\times 10^3$ fm$^3$, see figure 3.  The average number of
droplets nucleated is rather small, roughly 1/50 and 1/100 at
the AGS and SPS, respectively.  The reason for the smaller number
for set 2, SPS, is that the superheating is even more extreme,
and contributions from configurations other than spherical are
even more likely to be significant.

In figure 7 we plot the volume fraction converted to quark-gluon
plasma, $q(t)$, and the average droplet radius, $\bar{R}(t)$, defined as
\begin{equation}
\bar{R}(t) = \left[ \frac{3}{4\pi}\frac{q(t)}{n_{\rm drop}(t)}
\right]^{1/3} \, .
\label{rbar}
\end{equation}
Results for several values of the parameter $v_0$ are shown.
For $v_0 > 10$ we are in an asymptotic regime where both $q(t)$
and $\bar{R}(t)$ hardly change anymore from their values obtained
with $v_0 = 10$.  The maximum value for $q(t_f)$ is $9.2\times 10^{-3}$
and $3.2\times 10^{-3}$ for parameter set 1 and 2, respectively.
These values are somewhat disappointingly small.
On the other hand, the corresponding maximum average bubble radii
are 5.0 (4.3) fm for set 1 (2), which are interestingly large.

How are we to interpret these results?  If we were dealing with
an expansion chamber with a volume of 1 m$^3$ the answer would
be clear.  Droplets of quark-gluon plasma would be nucleated
randomly throughout the system.  Since the droplet density is
so small they would be widely separated.  They would grow to a
size of perhaps 3 to 5 fm and therefore would hardly ever touch
each other.  They would be scattered like stars in the night sky.
Only about 10$^{-3}$ to 10$^{-2}$ of the volume would be occupied
by plasma before the system cooled below the phase coexistence curve.
The interpretation for heavy ion collisions at the AGS, we claim,
is different.  By the end of the cooling period, at $t_f$, the
distribution of plasma droplets should be a Poisson function
in the variable $\bar{N}_{\rm drop} = V n_{\rm drop}$.
The likelihood that more than one droplet nucleates in a given
collision is very small.  Therefore, either one droplet nucleates
or none.  If one nucleates, it will have eaten a good fraction of
the hadronic matter, converting it to plasma.  Although we have
neglected feedback of plasma formation on the dynamical evolution
of the system, we can confidently say that plasma formation would
{\it slow} the expansion of the system.  This is due to the fact
that the pressure is much reduced when compared at the same
energy density; that is, the equation of state is softened by
the phase transition, and it is the pressure which drives the
expansion.  Therefore, the plasma has more time to eat the rest
of the hadronic matter.  Taking into account also the fact that
we have probably underestimated the nucleation rate, we conclude
that perhaps one in every 100 or 1000 central collisions at the
AGS will have undergone an almost complete phase transition
by the time the matter has expanded to the phase coexistence
curve.

\section{Discussion and Conclusions}

In this paper we have applied relativistic nucleation theory to
address the issue of quark-gluon plasma formation at the AGS and
at the lower end of the SPS energy range.  Our simple modeling
of the magnitude and time dependence of the baryon and energy
densities seem to be in reasonable agreement with those obtained in
hadronic cascade simulations.  The numerical results suggest that
perhaps one out of every 100 or 1000 central collisions will
exhibit a significant phase conversion from hadronic matter to
quark-gluon plasma.

Standard homogeneous nucleation theory assumes that the matter
undergoing the phase transition is not superheated or supercooled
too much from phase coexistence.  Central collisions of the most massive
nuclei at the AGS and the SPS apparently lead to {\it very} superheated
hadronic
matter.  As a consequence, the critical size plasma droplets have radii
which are comparable to or even smaller than the expected correlation
length.  Standard homogeneous nucleation theory also underestimates
the nucleation rate because configurations which are far from spherical
become important.  It may be that the superheating is so extreme that
one is approaching spinodal decomposition.

It is interesting to think about the transition from nucleation of
plasma at these relatively modest beam energies to the almost
instantaneous formation of plasma expected at RHIC and LHC energies.
In free space, hard nucleon-nucleon collisions produce a significant
number of secondary mesons.  It takes a finite time for the produced
quarks and gluons to actually hadronize as asymptotic meson states.
In a central nucleus-nucleus collisions these little star bursts
may overlap before hadronization can occur, thereby providing seeds
or nucleation sites for quark-gluon plasma.

Neither of the effects discussed above are taken into account in
our calculations.  Therefore, we have most likely {\it underestimated}
the formation of plasma at AGS and lower SPS energies, perhaps
significantly so.

What about experimental implications?  Since the phase transition
is occurring so far out of equilibrium we would expect a significant
increase in the entropy of the final state.  This could be seen in the
ratio of pions to baryons, for example, or in the ratio of deuterons
to protons \cite{me}.  Along with the increased entropy should come
a slowing down of the radial expansion due to a softening in the
matter, that is, a reduction in pressure for the same energy
density.  Together, these would imply a larger source size and
a longer lifetime as seen by hadron interferometry \cite{scott}.
Of course, these and other signals would have to be investigated
experimentally on an event by event basis.  It should be straightforward
to develop models of the probability distributions of the entropy to
baryon ratio, source size and lifetime, and so on.  Parameters
could be adjusted to learn about the probability of phase conversion
in a given central collision, the latent heat release, and so forth.
``Come, Watson, the game is afoot!" \cite{sh}

\section*{Acknowledgements}

We thank J. Sandweiss for a stimulating conversation.
This work was supported by the U.S. Department of Energy under
Grant No. DOE/DE-FG02-87ER40328.

\newpage
\section*{Appendix}

The classical theory of nucleation culminated in the work of
Becker and D\"oring \cite{Becker}.  It is nicely reviewed by
McDonald \cite{Mac}.  It was developed to describe the nucleation
of a liquid droplet in a dilute yet supersaturated vapor.  Langer's
formalism, as used in this paper, is meant to apply when neither
phase may be considered dilute.  We do not believe that the superheated
hadronic matter is dilute enough to apply the classical theory.
Nevertheless, for comparison we would like to summarize the
nucleation rate in the classical regime.

The classical expression for the nucleation of a droplet of dense
liquid in a dilute gas is
\begin{displaymath}
I = a(i_*) \left(\frac{|\Delta E''(i_*)|}{2\pi T}\right)^{1/2}
n_1 \exp{\left( -\Delta E(i_*)/T \right)} \, ,
\end{displaymath}
where $\Delta E(i_*)$ is the formation energy of a critical sized droplet
consisting of $i_*$ molecules, prime denotes differentiation with respect
to the number of molecules $i$, T is the temperature, $n_1$ is the
density of single molecules and $a(i_*)$ is the accretion rate of single
molecules on a critical droplet.  Usually the accretion rate is taken
to be
\begin{displaymath}
a(i_*) \,=\, \frac{1}{2} n_1 \bar{v} 4\pi R_*^2 s \, ,
\end{displaymath}
which is the flux of particles ($\bar{v}$ is the mean speed of gas
molecules) striking the surface of the critical droplet times
a `sticking fraction' $s$ less than one.  The first term in the
nucleation rate is a dynamical factor influencing the growth rate, the
second term characterizes fluctuations about the critical droplet, and
the product of the third and fourth terms gives the quasi-equilibrium
number density of critical sized droplets.  The energy is measured
with respect to the gas molecules so that $\Delta E(1) = 0$.

To attempt to apply the classical expression to the nucleation of
a plasma droplet, the first thing we do is to multiply the
Boltzmann factor by the number of states available to the hot
droplet.
\begin{displaymath}
e^{-\Delta E/T} \rightarrow e^{-\Delta E/T} e^{\Delta S}
\end{displaymath}
Due to the thermodynamic identities $S = -dF/dT$ and $F = E - TS$
this modifies the Boltzmann factor to $e^{-\Delta F/T}$.

We prefer to characterize the size of the droplet not by the number
of molecules it contains but by its radius.  Then integration over
quadratic fluctuations about the mean size will give the prefactor
\begin{displaymath}
\left( \frac{|\Delta F''(R_*)|}{2\pi T} \right)^{1/2}
\end{displaymath}
The accretion rate must be multiplied by the increase in radius
per particle absorbed to compensate for this change of variable.
Upon absorption of one more particle the droplet free energy
changes by
\begin{displaymath}
\delta \Delta F \,=\, \Delta F'(R_*) \delta R
+\frac{1}{2} \Delta F''(R_*) \left( \delta R \right)^2 \, .
\end{displaymath}
The derivatives are evaluated at $R_*$ whereupon the first derivative
vanishes.  The (Gibbs) free energy added by one
gas molecule is just minus the pressure of the gas molecules divided
by their number density.  Therefore the accretion rate is multiplied
by the factor
\begin{displaymath}
\delta R \,=\, \left( -\frac{P_1}{n_1 \Delta F''(R_*)} \right)
^{1/2} \, .
\end{displaymath}
Putting everything together we arrive at
\begin{displaymath}
I \,=\, 2\pi s \bar{v} R_*^2 n_1^2 \left( \frac{P_1}{n_1\pi T} \right)^{1/2}
\exp{\left( -\Delta F_*/T \right)} \, .
\end{displaymath}
Generalizing to different species of molecules (hadrons) we write
\begin{displaymath}
I \,=\, 2\pi R_*^2 n_0 \exp{\left(-\Delta F_*/T \right)}
\, \Sigma_j s_j \bar{v_j}
n_j \left( \frac{P_j}{n_j \pi T} \right)^{1/2} \, ,
\end{displaymath}
where $P_j$ is the partial pressure of the $j$'th species, $n_j$
is their density, etc.  The quasi-equilibrium density of critical
droplets is normalized to the density of the lightest species of
hadrons, $n_0$.

Note especially the appearance of $R_*^2$ in the prefactor.  This
arises from the fact that the absorption rate is proportional to
the surface area.  In contrast, when the growth rate is dominated
by dissipation the prefactor has only one power of $R_*$.
Over most of the cooling curve it turns out that the prefactor
estimated in this classical approach is about the same order of
magnitude as the prefactor used in the text.

\newpage

\section*{Table}

\begin{center}

\begin{tabular}{|c|c|c|}\hline
\,& set 1 & set 2 \\ \hline
$E_{\rm beam}$ (GeV) & $11.6$ & $25.0$ \\ \hline
$A$ & $197$ & $208$ \\ \hline
$\gamma$ & $2.68$ & $3.78$ \\ \hline
$t_0$ (fm/c) & $2.56$ & $1.85$ \\ \hline
$t_f$ (fm/c) & $7.0$ & $6.95$ \\ \hline
$T$ (MeV) & $173$ & $214$ \\ \hline
$\mu_B$ (GeV) & $1.72$ & $2.06$ \\ \hline
$P_{\rm had}$ (GeV/fm$^3$) & $0.67$ & $1.41$ \\ \hline
$P_{\rm qg}$ (GeV/fm$^3$) & $2.19$ & $5.07$ \\ \hline
$\varepsilon_{\rm had}$ (GeV/fm$^3$) & $1.95$ & $3.90$ \\ \hline
$\varepsilon_{\rm qg}$ (GeV/fm$^3$) & $7.78$ & $16.44$ \\ \hline
$n_B^{\rm had}$ (fm$^{-3}$) & $0.78$ & $1.10$ \\ \hline
$n_B^{\rm qg}$ (fm$^{-3}$) & $3.17$ & $5.58$ \\ \hline
$s_{\rm had}$ (fm$^{-3}$) & $7.42$ & $14.27$ \\ \hline
$s_{\rm qg}$ (fm$^{-3}$) & $25.93$ & $46.94$ \\ \hline
\end{tabular}
\end{center}

\noindent Table 1 : Initial conditions for the two chosen parameter sets and
some resulting characteristic scales.

\newpage

\section*{Figure Captions}

Figure 1: Phase diagram of strongly interacting matter in the
temperature - baryon chemical potential plane (top panel)
and in the temperature - baryon density plane (bottom panel).
The dashed curve represents phase coexistence between
hadronic and quark-gluon matter.  It does not extend to zero
temperature because our description is too crude there.  The solid
curves represent the trajectories followed by heavy ion
collisions (neglecting nucleation of plasma) for parameter sets 1
and 2 in our simplified model.\\

\noindent Figure 2: The free energy difference $\Delta F(R)$ between a
hadronic phase with and without a quark--gluon plasma droplet.
This corresponds to the starting point for a collision at the
AGS, parameter set 1.\\

\noindent Figure 3: Time evolution of the volume $V(t)$ of the
collision region.\\

\noindent Figure 4: Upper row: Time evolution of the baryon chemical
potential $\mu_B$ and the temperature $T$ for parameter set 1 (2)
on the left (right).
Lower row: Time evolution of the baryon number density $n_B^{\rm had}$
and the energy
density $\varepsilon_{had}$ for parameter set 1 (2) on the left (right).\\

\noindent Figure 5: Time evolution of the nucleation rate $I(t)$
along the dynamical trajectories for both parameter sets.\\

\noindent Figure 6: Time evolution of the average plasma droplet
number density $n_{\rm drop}$ for both parameter sets.\\

\noindent Figure 7: Upper row: Time evolution of the quark--gluon
fraction $q$ for different values of $v_0$ for parameter set
1 (2) on the left (right).
Lower row: Time evolution of the average droplet radius $\bar{R}$ for
different values of $v_0$ for parameter set 1 (2) on the left (right).


\begin{thebibliography}{99}
\bibitem{note1} By charge symmetry we mean that the ratio of
electric charge to baryon number is 1/2.  This ratio of actually
somewhat less than 1/2 for heavy nuclei like gold or lead.
We are not concerned with such fine details in this paper.
\end{thebibliography}
\end{document}